# Electric gating induced bandgaps and enhanced Seebeck effect in zigzag bilayer graphene ribbons

Thanh-Tra Vu [1), Van-Truong Tran [2)

[1) Physics Department, School of Education, Can Tho University, Can Tho, Vietnam

[2) IEF, Université Paris-sud, CNRS, UMR 8622, Bât 220, 91405 Orsay, France.

Emails: vttra@ctu.edu.vn and van-truong.tran@u-psud.fr

**Abstract**

We theoretically investigate effect of a transverse electric field generated by side gates and a vertical electric field generated by top/back gates on energy bands and transport properties of zigzag bilayer graphene ribbons (Bernal stacking). Using atomistic Tight Binding calculations and Green's function formalism we demonstrate that bandgap is opened when either field is applied and even enlarged under simultaneous influence of the two fields. Interestingly, although vertical electric fields are widely used to control bandgap in bilayer graphene, here we show that transverse fields exhibit more positive effect in terms of modulating a larger range of bandgap and retaining good electrical conductance. Seebeck effect is also demonstrated to be enhanced strongly about 13 times for a zigzag bilayer graphene ribbons with 16 chain lines. These results may motivate new designs of devices made of bilayer graphene ribbons using electric gates.

# 1  Introduction

Since the experimental discovery by Geim and Novoselov,[1] Graphene has become an attractive topic to researchers due to its appealing electronic properties with linear energy dispersion around neutrality point and extremely high electron mobility.[2] Inspired of huge potentials for electronic applications, graphene suffers from an obstacle of lacking an energy gap, which made it indirect suitably for many applications, especially for transistors.[3–5] Different strategies of nanostructure design have been taken to open bandgap in mono layer graphene, such as cutting graphene sheet into ribbons and bandgap can be observed thank to the fine size effect[6] or introducing nano holes in graphene sheets to open mini gaps.[7]

Other strategies like doping outer atoms, for instance, doping Nitrogen or Boron atoms into graphene sheet [8,9] or heterostructures of graphene and a similar structure material such as Boron Nitride [10,11] have also been applied recently.

Another strategy to open an energy bandgap in graphene is to use bilayer structures and put them in an external electric field. This approach was first studied theoretically by McCann and Fal'ko in 2006 [12] and it was confirmed experimentally by Ohta et. al.[13] in the same year. All studies on the effect of a vertical electric field in 2D bilayer graphene showed that maximum bandgap can be achieved is about 250 meV.[14–17]

The effect of vertical field was found similar in metallic bilayer graphene nanoribbons (BGNRs) [18,19] while it behaves differently in semiconducting BGNRs with a reduction of bandgap.[20] In case of single layer graphene nanoribbons (SGNRs), transverse electric fields was used to modulate bandgap.[21–24] It has been demonstrated that by applying a transverse field, bandgap can be open in case of metallic armchair and zigzag ribbons[21] while it can be strongly suppressed in case of finite bandgap ribbons.[21,24] Recently, ideas of using transverse field to modulate bandgap and then controlling current in electronic devices have been applied successfully.[24,25]

It is thus in the push for practical applications, it is desirable to have the ability to modulate the graphene structures functionalities by an external stimulus. In this communication, we propose a generic approach by using electric gates to control of the intriguing properties in zigzag bilayer graphene nanoribbons. By using Tight Binding (TB) calculations and combining both transverse in-plane and vertical electric fields, our results show that bandgap in zigzag BGNRs with A-B stacking even can be tuned more flexibly compared to other studies using only one field. The bandgap is found to reach its largest value under simultaneous effect of the two fields. Employing Green's function for transport study, surprisingly results exhibit that transverse field is more effective than vertical one since it induces a larger gap and preserves good electrical conductance. Seebeck effect also has a significant enhancement thanks to the opening of bandgap.

## 2 Modeling and methodologies

## 2.1 Studied structures

In order to understand effect of different fields on energy bands and transport properties of zigzag BGNRs, we consider a structure made of a zigzag BGNR under effect of external fields as designed in figure 1. The top (red) and bottom (blue) layers are referred to as layer 1 and layer 2, respectively. The width of each sub-ribbon is characterized by the number of chain lines *M*. Figure 1(a) is the schematic view of the system with two side gates are placed parallel with the edges of the ribbons. These two gates are applied by two potential + $V_s$/2 and – $V_s$/2 to generate a transverse electric field $\vec{E}_\parallel$ with the average strength $E_\parallel = V_s / W$, where *W* is the width of the zigzag BGNR. A sketch of a vertical field $\vec{E}_\perp$ generated by top + $V_t$/2 and back – $V_t$/2 gates is also shown in figure 1(b). In figure 1(c), intra-layer and inter-layer couplings between atoms are illustrated, $t_0$ is hoping energy between two nearest atoms in a single layer, meanwhile $t_1$, $t_3$ and $t_4$ are hoping energies between nearest atoms (A-B), next-nearest atoms (A-B) and next-nearest atoms (A-A, B-B) in two layers, respectively.[26–28]

For energy analysis, bilayer ribbons are assumed to be infinite and under the effect of electric fields. And for transport study, two leads made of zigzag BGNRs are connected to the active regions as sketched in figures 1(a) and 1 (b).

## 2.2 Methodologies

Electronic properties of zigzag BGNRs are examined by using TB model. The general form of Hamiltonian can be written simply as [29]

$$H = \sum_i U_i |i\rangle\langle i| - \sum_{\langle i,j \rangle} t_{ij} |i\rangle\langle j|, \tag{1}$$

where $U_i$ is total energy potential at *i*-th site, $U_i = -e.(V_t/2 + V_i)$ if *i*-th site belongs to layer 1 and $U_i = -e.(-V_t/2 + V_i)$ if this site belongs to layer 2. $V_i = -V_s/2 + E_\parallel . y_i$ is the electrostatic potential at *i*-th site due to the transverse field and $y_i$ is the position of the site with respect to the side gate – $V_s$/2. $t_{ij}$ is the coupling between atoms at *i*-th site and its surrounded neighbor atoms. In single layer, only nearest neighbor interactions are taken into account, so $t_{ij} = t_0$. Regarding to interlayer interactions between the two layers, $t_{ij}$ can fit to $t_1$, $t_3$ or $t_4$ depend on distance between atoms *i*-th and *j*-th. In this works, hoping parameters were taken from ref. [26,27], with $t_0$ = 2.598 eV, $t_1$ = 0.364 eV, $t_3$ = 0.319 eV and $t_4$ = 0.177 eV.

Relaxation effect at the edges of each sub-ribbon is also considered. It was reported in ref. [22] that to fit with ab initio calculations, the tight binding hoping parameter at the edges of SGNRs must be increased about $\delta_t = 12\%$. This effect on electronic properties of zigzag BGNRs will be discussed later.

To investigate transport properties, the non-equilibrium Green's function (NEGF) approach was employed within the ballistic approximation.[30–32] In this formalism, the transmission $T(E)$ is computed as $T = \text{Trace}(\Gamma_L G \Gamma_R G^\dagger)$, where $G = [E - H_D - \Sigma_L - \Sigma_R]^{-1}$ is the retarded Green's function of the active region, and $\Gamma_{L(R)} = i(\Sigma_{L(R)} - \Sigma^\dagger_{L(R)})$ is the injection rate at the interfaces of the left (right) leads and the active region. Transmission then is used to calculate the intermediate function $L_n(\mu, T)$ which is defined as [33]

$$L_n(\mu, T) = \frac{2}{h} \int_{-\infty}^{+\infty} dE\, T_e(E)(E - \mu)^n \frac{-\partial f_e(E, \mu, T)}{\partial E}, \qquad (2)$$

Where $f_e(E, \mu, T)$ is the Fermi distribution function, *T* is the absolute temperature and *μ* is the electron chemical potential. Electrical conductance and Seebeck coefficient are computed via the inter-mediate functions by Landauer's-Onsager's approach[33]

$$\begin{cases} G_e(\mu, T) = e^2 L_0(\mu, T) \\ S(\mu, T) = \frac{1}{eT} \frac{L_1(\mu, T)}{L_0(\mu, T)} \end{cases} \qquad (3)$$

## 3  Results and discussion

In figure 2, we show energy bands (black solid lines) of a zigzag BGNR for *M* = 16 chain lines for different potentials of top/back and side gates applied. Figure 2(a) is the band structure without external fields and the obtained bands are quite similar to those of a zigzag SGNR but with twice numbers of states at the same energy level. Flat bands appear near neutrality point are edge states. The inset indicates that there is a small bandgap about 15 meV in this structure. The bandgap becomes visible when a side or top/back potential applied individually as shown in figures 2(b) and 2(c). Notably, with the same strength of potentials, the side gates potential seems to have stronger influence on bandgap than the top/back gates potential, i.e., a bandgap is about 310 meV obtained for $V_s = 0.5$ V as shown in figures 2(b) and 184 meV for $V_t = 0.5$ V is applied as shown in figures 2(c). Interestingly, if we applied two fields simultaneously with $V_t = 0.5$ V and $V_s = 0.5$ V, the bandgap is strongly suppressed and almost equal to zero as seen in figure 2(d).

To understand the effect of each individual field and their total effect on energy bands of the zigzag BGNRs we plot the density of state (DOS) and show it on the right side of each energy structure in figure 2. Moreover, in order to understand effect of these fields on each layer of the bi-layer ribbon, we separate the contributions of each layer in the total density of state (TDOS) by plotting the local density of state (LDOS) in each layer. In all panels of DOSs, the dash blue and dot black curves are LDOSs in layer 1 and layer 2, respectively. In the meanwhile the solid cyan curves are TDOS.

First of all, in figure 2(a) we observe high sharp peaks around zero energy in TDOS due to flat bands of edge states. However, unlike the case of zigzag SGNRs, in this case of zigzag BGNRs there are two peaks near zero energy instead of one. These two peaks also appear in LDOS of each layer as seen in the dash blue and dot black curves. Because the peak of TDOS near neutrality point was proved to be corresponding to edge states at the edges of a zigzag SGNR.[34,35] These two peaks indicate that edge states at two edges of each sub-ribbon of a BGNR are separated and not degenerated as in a zigzag SGNR.[34,35] This result is due to the fact that second neighbor interlayer interactions $t_3$ and $t_4$ were taken into calculations. Physically, it leads to a broken symmetry between two edges in each sub-ribbon due to number of interactions at the two edges are different. This characteristic can be also seen in figure 1(a).

With the appearance of a transverse electric field generated by two side gates, we observe a dramatically change in DOS spectrum in figure 2(b), especially in low energy region. Zero DOS around the neutrality point reflects the open of an energy gap and separates peaks indicating the shift of edge states. LDOS of each layer shows that under the effect of transverse field, in both layers, in particular at low energy region, conduction states move up and valence states move down and finally it results in an opening of bandgap. In the other hand, LDOSs of layer 1 and layer 2 also indicate that the effects of transverse field on two layers are almost the same everywhere exception the region near the neutrality point. This result can be understood by noting the relative position of each layer with respect to the side gates. In figure 1(a) we can see that one edge of layer 1 is closer to the gate – $V_s/2$ than the counterpart of layer 2. In contrast, at the other side, the edge of layer 2 is closer to the gate + $V_s/2$ compared to the other of layer 1. In other words, the gate – $V_s/2$ has a little more influence on layer 1 meanwhile the gate + $V_s/2$ supports more for layer 2. This effect leads to a moderate difference of the effect of the field on the two layers.

The field effect on each layer is much intense in the case of a perpendicular field which is generated by top and back gates. In figure 2(c), LDOSs of layer 1 and layer 2 are totally separated at any range of energies. In comparison with LDOS without fields applied in figure 2(a), we can see that the vertical field shifts down all bands of layer 1 while it shifts up all bands of layer 2. Finally a bandgap is open and it is determined by the lowest conduction band of layer 1 and the highest valence band of layer 2.

It is thus the openings of bandgap by the transverse and vertical fields are based on different mechanisms. In figure 2(d) when two fields are applied simultaneously with the same strength, two highest peaks in LDOS of layer 1 or layer 2 show that edge states at two edges of each sub-ribbon are separated compared to the case without fields applied in figure 2(a). This result is due to the effect of the transverse field as discussed above. However, these two peaks are not localized in two opposite sides of the neutrality point as in figure 2(b), but they are shifted down in layer 1 and shifted up in layer 2. This is consequence of the effect of the vertical field. Finally, edge states of one edge of layer 1 (layer 2) localized at zero energy and the bandgap is close. Beyond this analysis we can predict that zero bandgap is likely observed in case $V_t = -0.5$ V and $V_s = 0.5$ V because in this case the vertical field will shift up states in layer 1 and shift down states in layer 2, but eventually there are still two edge states peaks at zero energy that each one contributed by one layer.

To go further into the analysis of the total effect of these two studied fields we plot energy gap $E_{gap}$ as a function of $V_s$ and $V_t$ and show in figure 3. In figure 3(a) we can observe that the effect of the external fields can be classified into four regions which are confined by the cones $V_s = V_t$ and $V_s = -V_t$. On the edges of these cones, the bandgap is almost zero as it can be seen in figure 2(d). Interestingly, the largest bandgap can be obtained in the region 1 or 3 of ($V_s$, $V_t$), i.e., $E_{gap}$ is maxima at $V_s = -1.1$ V, $V_t = 0.15$ V or $V_s = 1.1$ V and $V_t = -0.15$ V. Similar results are obtained for other widths of bilayer ribbons, for instance, $M = 13$ as shown in figure 3(b). It is worth to note that the color in figure 3(a) and (b) indicates that the largest bandgap is dependent on the width $M$ of BGNRs.

In figure 4 we plot the effect of individual field (figures 4(a) and 4(b)) and the mutual effect on the bandgap (figure 4(c)) for different widths of zigzag BGNRs. In figure 4(a) we can observe that exception for the region near $V_s = 0$, basically $E_{gap}$ increases with the increase of $V_s$, then it reaches to maximum value before falling down. The maximum value of $E_{gap}$ and its position is strongly dependent on the width of sub-ribbons, i.e., for $M = 13$, 16 and 19 we obtain maximum of energy gap max($E_{gap}$) is about 560 meV, 450 meV and 369 meV, respectively. The position $V_s$ of the peak is shifted close to $V_s = 0$ when $M$ increases, for instance, max($E_{gap}$) is at $V_s = \pm 1.44$ V for $M = 13$, and it is at $V_s = \pm 1.14$ V, $\pm 0.92$ V for $M = 16$ and 19, respectively. This behavior of transverse field is similar to that obtained in zigzag SGNRs.[21] For the case $V_t$ is varied and $V_s$ is fixed equal to 0, a similar physics is obtained.

In case $V_t = 0.15$ V is applied meanwhile $V_s$ is varied from $-1.5$ V to 1.5 V (figure 4(c)), first, $E_{gap}$ reduces from 60 meV to zero and then it grows up to the maximum value as since in case $V_t = 0$ in figure 4(a). In this case we do not see the symmetry between $+ V_s$ and $- V_s$ because of the additional effect of $V_t$. In fact, the bandgap seems to be enlarged in the region $- V_s$ and narrowed in the region $+ V_s$. This outcome can be seen again in figure 3. In figure 4(d), max($E_{gap}$) is considered when varying fields for different ribbon widths $M$. The open circle blue line is the case no field is applied, the bandgaps of BGNRs are close to zero (about 10-15 meV). The diamond green line refers to the maximum of bandgaps for $V_s = 0$, $V_t$ varies from $-1.5$ V to 1.5 V and the square red line stands for the case $V_t = 0$, $V_s$ is in [$-1.5$ V, 1.5 V]. Comparing the red and green lines we can see that the transverse field is more efficient than the vertical field in term of modulating large range of bandgap. For example, for $M = 16$ max($E_{gap}$) equals to 271 meV when $V_t$ varies, while it equals to 450 meV with the change of $V_s$. The triangular purple line indicates that the bandgap is enlarged when varying both $V_s = [-1.5$ V, 1.5 V], $V_t = [-1.5$ V, 1.5 V]. The maximum of the bandgap increases about 10.7% for $M = 16$ and 21% for $M = 25$. The filled circle black line is plotted for the case single layer zigzag graphene ribbons are under a transverse field with $V_s$ varies in [$-1.5$ V, 1.5 V]. It is thus the maximum bandgap obtained in zigzag BGNRs is not exceeded than that obtained in zigzag SGNRs. However, except for the filled circle dashed back line for the case of zigzag SGNRs (also $M = 16$), other results in figure 5(a) are for zigzag BGNRs and it indicates that the maximum bandgap of the zigzag BGNR is similar to that of the zigzag SGNR but it has much higher electrical conductance. Hence, under the effect of electric fields, it is expected that on/off ratio of current in devices made of zigzag BGNRs is higher than that in devices made of zigzag SGNRs.

The solid red line is the conductance without external fields. We also plot the open circle black line for the case of perfect ribbons with no edge relaxation for comparison. It can be seen clearly that the effect of edge relaxation is weak. This observation agrees with the conclusion on zigzag SGNRs reported in ref [22]. However this effect is more significant for armchair ribbons.[20,22] The dashed blue line is the conductance for case $V_t = 0$ and $V_s = 0.8$ V and the dot green one is for case $V_t = 0.8$ V and $V_s = 0$. The results show a surprise that the conductance with the vertical field applied is smaller than that of the counterpart field in most range of chemical energy. This behavior can be understood by comparing energy bands and DOSs in figures 2(b) and 2(c) with those in figure 2(a). It can be seen that the vertical field has stronger influence on energy bands and DOSs than the transverse field, which thus resulting a stronger electron scattering in the active region of the device and conductance must be smaller. When two fields applied simultaneously with $V_t = 0.15$ V and $V_s = -1.1$ V (the dashed dot cyan line), the conductance stays between the dashed blue and the dot green lines.

In the case of 2D graphene, Seebeck coefficient is less than 100 µV/K due to the lack of a gap.[36] Thank to the opening of bandgap, Seebeck effect is likely enhanced in the structures studied here because of the separation of electrons and holes.[37–39] In figure 5(b) we plot Seebeck coefficient for different cases of external fields applied. The inset with solid red curve shows the peak of Seebeck coefficient $S$ reaches value 54 µV/K for the case no field is applied. It is thus $S$ increases about 6.5 times when $V_t = 0.8$ V, $V_s = 0$ and raises to 11.3 times when $V_t = 0$, $V_s = 0.8$ V. Seebeck coefficient even reach 719 µV/K, about 13.3 times higher when the bandgap gets the maximum with $V_t = 0.15$ V, $V_s = -1.1$ V. Because modulated bandgap is dependent on the ribbon width as shown in figure 4(d), the value of $S$ will increase lager for $M < 16$ and reduce smaller for $M > 16$.

## 4  Conclusion

We have investigated field effect to electronic properties of bilayer zigzag graphene ribbons. Both transverse and vertical fields are considered. Though the vertical fields have been widely used to open bandgap in 2D bilayer graphene, it has shown that transverse field generated by side gates can induce larger gap than vertical field induced by top/back gates. Moreover, the investigation on electrical conductance indicated that transverse field causes less electron scattering than that by vertical field, leading to a good electrical conductance. In the case two fields are applied simultaneously, bandgap can be enlarged up to 21% for $M = 25$. Seebeck effect has been also investigated and it has shown that Seebeck coefficient increases from about 54 uV/K without fields applied to 719 uV/K with fields induced largest bandgap. This effective doping effect due to the types of fields (transverse and vertical) is tremendously important for understanding how the electric gates can modulate the electronic structures. Moreover, this study may open a new avenue to design and engineer new functional devices made of bilayer graphene ribbons by using electric gates.

## Acknowledgments

The authors acknowledge Vietnam's National Foundation for Science and Technology Development (NAFOSTED) for financial support through the project no. 103.01-2015.98.

nanoribbons by graphene/BN interface engineering *Nanotechnology* **26** 495202

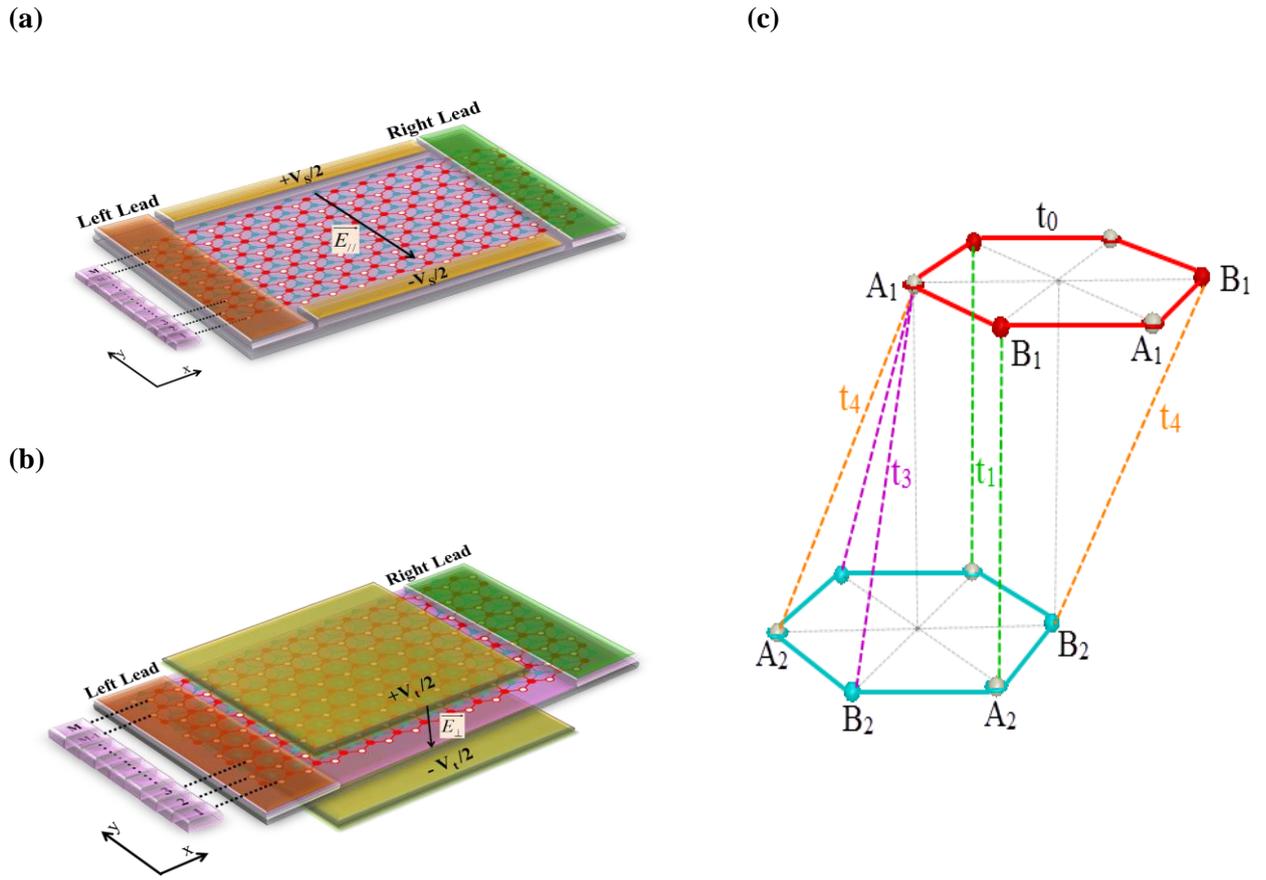

Fig 1: Schematic view of zigzag bilayer graphene ribbons (AB stacking) under the effect of (a) a transverse electric field generated by two side gates $+V_s/2$ and $-V_s/2$ (b) a vertical electric field generated by a top gate $+V_t/2$ and a back gate $-V_t/2$. (c) Illustration of intra-layer and inter-layer interactions between atoms in a single layer and two layers.

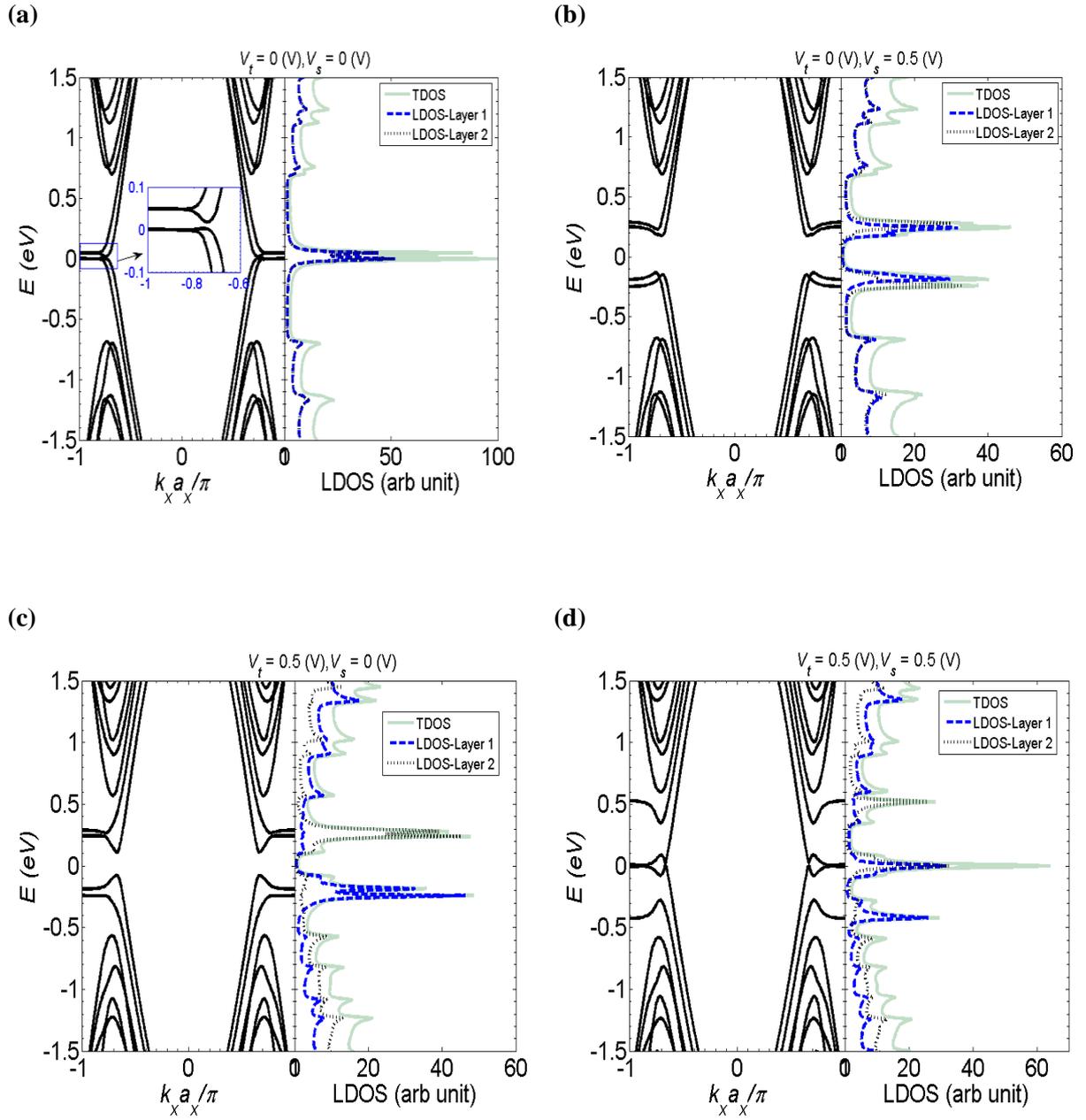

Fig 2: Energy bands and the local density of states (LDOS) for different potentials applied: (a) No field is applied, (b) only transverse field is applied for $V_s = 0.5$ V, (c) only vertical field is applied for $V_t = 0.5$ V and (d) two fields are applied simultaneously for $V_t = 0.5$ V, $V_s = 0.5$ V.

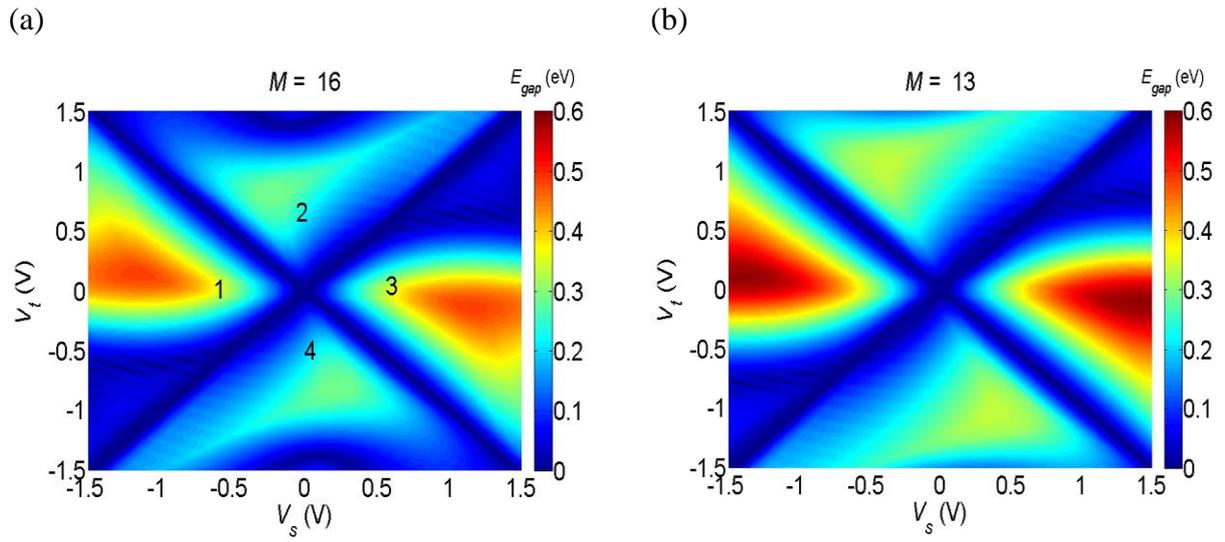

Fig 3: Bandgap is plotted as a function of $V_t$ and $V_s$ for different widths of ribbons (a) $M = 16$ and (b) $M = 13$.

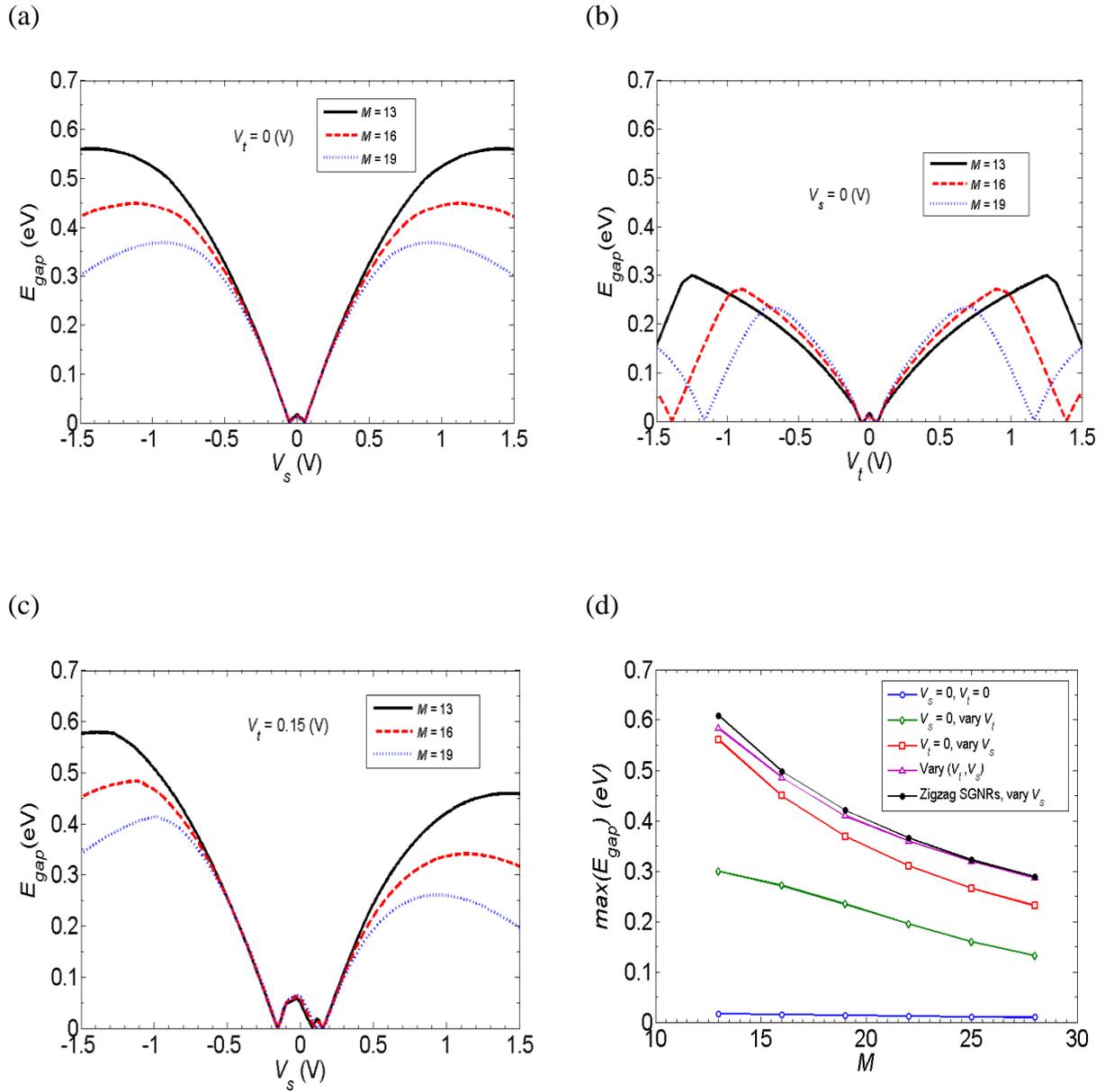

Fig 4: Bandgap is plotted as a function of (a), (c) $V_s$ or (b) $V_t$ for a fixed value of its counterpart $V_t$ or $V_s$. (d) Maximum of $E_{gap}$ is considered as a function of ribbon width for different cases of fields varying from – 1.5 V to + 1.5 V.

(a) (b)

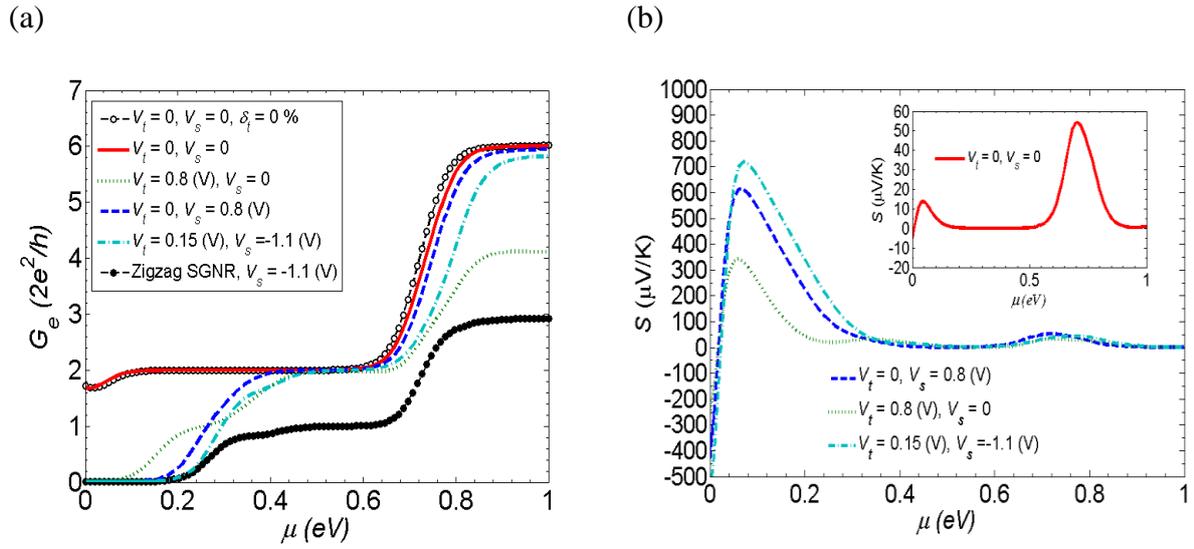

Fig 5: (a) Conductance and (b) Seebeck coefficient are plotted as functions of chemical energy for different values of the couple ($V_t$, $V_s$). Inset in (b) is Seebeck coefficient without fields applied. Here $M = 16$, the length of the active region is about 25.4 nm (60 unit cells) and simulations are performed at room temperature $T = 300$ K.